\documentclass[twocolumn]{emulateapj}  
\usepackage{epstopdf}
\usepackage{ulem}
\usepackage{amssymb}
\usepackage{natbib}
\usepackage{times}
\usepackage{graphicx,bm,amssymb}
\usepackage{pstricks}
\usepackage{psfrag}
\usepackage[colorlinks=true,linkcolor=blue,citecolor=blue]{hyperref}
\newcommand{\be}{\begin{equation}}
\newcommand{\ee}{\end{equation}}
\newcommand{\bse}{\begin{subequations}}
\newcommand{\ese}{\end{subequations}}
\newcommand{\bary}{\begin{eqnarray}}
\newcommand{\eary}{\end{eqnarray}}

\def\aj{AJ}
\def\araa{ARA\&A}
\def\apj{ApJ}
\def\apjl{ApJ}
\def\apjs{ApJS}
\def\aap{A\&A}
\def\mnras{MNRAS}
\def\na{New A}
\def\nat{Nature}
\def\physrep{Phys.~Rep.}

\shorttitle{THEORETICAL DESCRIPTION OF GRB 160625B}
\shortauthors{Fraija N. et al.}
\begin{document}
\title{THEORETICAL DESCRIPTION OF GRB 160625B WITH WIND-TO-ISM TRANSITION AND IMPLICATIONS FOR A MAGNETIZED OUTFLOW}
\author{N. Fraija$^{1\dagger}$, P. Veres$^2$,   B. B. Zhang$^{3}$,  R. Barniol Duran$^4$,  R. L. Becerra$^1$,  B. Zhang$^5$,   W. H. Lee$^1$,  A. M. Watson$^1$, C. Ordaz-Salazar$^1$ and A. Galvan-Gamez$^1$} 
\affil{$^1$ Instituto de Astronom\' ia, Universidad Nacional Aut\'onoma de M\'exico, Circuito Exterior, C.U., A. Postal 70-264, 04510 M\'exico D.F., M\'exico.\\
$^2$ Center for Space Plasma and Aeronomic Research (CSPAR), University of Alabama in Huntsville, Huntsville, AL 35899, USA\\
$^3$ Instituto de Astrof\'isica de Andaluc\'ia (IAA-CSIC), P.O. Box 03004, E-18080 Granada, Spain\\
$^4$ Department of Physics and Astronomy, Purdue University, 525 Northwestern Avenue, West Lafayette, IN 47907, USA\\
$^5$ Department of Physics and Astronomy, University of Nevada Las Vegas, Las Vegas, NV 89154, USA\\
}
 \email{$\dagger$nifraija@astro.unam.mx}
%
\begin{abstract}
GRB 160625B, one of the brightest bursts  in recent years, was simultaneously observed by Fermi and Swift  satellites, and ground-based optical telescopes in three different events separated by long periods of time.   In this paper the non-thermal multiwavelength observations of GRB 160625B are described and  a  transition phase from wind-type-like medium to interstellar medium between the early (event II) and the late (event III) afterglow  is found.  The multiwavelength observations of the early afterglow are consistent with the afterglow evolution starting at $\sim$ 150 s in a stellar wind medium whereas the observations of the late afterglow  are consistent with the afterglow evolution in interstellar medium (ISM).  The wind-to-ISM transition is calculated to be at $\sim 8\times 10^3$  s when the jet has decelerated, at a distance of $\sim$ 1 pc from the progenitor.   Using the standard external shock model, the synchrotron and synchrotron self-Compton emission from reverse shock is required to model the GeV $\gamma$-ray and optical  observations in the  early afterglow, and  synchrotron radiation from the adiabatic forward shock to describe the X-ray and optical observations in the late afterglow.  The derived values of the magnetization parameter,  the slope of the fast decay of the optical flash and the inferred magnetic fields suggest that  Poynting flux-dominated jet models with arbitrary magnetization could account for the spectral properties exhibited by GRB 160625B.
 \end{abstract}
\section{Introduction}
Gamma-ray bursts (GRBs) are the most luminous explosions in the universe.  Observations have firmly established that GRB prompt phases and their afterglows arise from highly relativistic and collimated outflows \citep{2002ApJ...571..779P, 2004ApJ...609L...1T, 2015PhR...561....1K}. Long GRBs (lGRBs) have been associated  to the core collapse of massive stars \citep{2006ARA&A..44..507W, 2012grbu.book..169H, 2003Natur.423..847H}. According to the collapsar model, lGRBs are generated in shocks that take place after the ultra relativistic jet has broken out from the stellar envelope. The jet dynamics is mainly dominated by the jet head, which is controlled by the difference in pressures between the reverse and forward shocks.  If the luminosity is low enough and/or the density of stellar envelope is high enough, the collimated jet  will then be surrounded by a cocoon  \citep{2013ApJ...777..162M,2011ApJ...740..100B, 2002MNRAS.337.1349R}.  When the relativistic jet is going through the progenitor star, its rate of advance is slowed down and most of the energy output during this phase is deposited into the cocoon. It starts spreading up to the optical depth becomes equal to unity and then, an X-thermal component could be expected.\\
The description of bright optical flashes by reverse shocks  \citep{2015ApJ...810..160G, 2016ApJ...818..190F, 2014Sci...343...38V, 2016ApJ...833..100H,  2003ApJ...595..950Z} and the high degree of optical polarization detected in some bursts \citep{2016MNRAS.455.3312G, 2014NewA...29...65P, 2015ApJ...813....1K, 2009Natur.462..767S, 2007Sci...315.1822M, 2017Natur.547..425T, 2013Natur.504..119M} have supplied strong evidence that sources could be endowed with magnetic fields \citep{1992Natur.357..472U, 2003Natur.423..415C, 2000ApJ...537..810W}.  Using standard assumptions such as  the reverse-shocked shell carries a substantial energy,  optical flashes are described by synchrotron emission from reverse shock which is shown as a single peak \citep{2000ApJ...536..195C, 2005ApJ...628..315Z, 2003ApJ...595..950Z, 2003ApJ...597..455K, 2000ApJ...545..807K} and then high-energy photons could be generated by inverse Compton scattering process \citep{2001ApJ...546L..33W, 2001ApJ...556.1010W, 2007ApJ...655..391K, 2015ApJ...804..105F, 2016ApJ...831...22F}. \\
\cite{2013ApJS..209...11A} reported the first Fermi-LAT catalog which summarized the temporal and spectral properties of the 28 GRB LAT-detected above 100 MeV and 7 GRBs above $\sim$ 20 MeV.  These bursts were recorded since the beginning of nominal science operations in 2008 until 2011.  Analysis of the high-energy emission showed that the more luminous bursts present a bright and short-lasting peak at the end of the prompt emission and  a temporally extended component lasting hundreds of seconds.\\
GRB160625B was detected on 2016 June 25 by both instruments on board Fermi satellite; Gamma-Ray Burst Monitor \citep[GBM;][]{2016GCN..19581...1B} and Large Area Telescope \citep[LAT;][]{2016GCN..19586...1D},  XRT and UVOT instruments on board Swift satellite \citep{2016GCN..19585...1M} and several optical telescopes \citep[CASANDRA all-sky cameras on the BOOTES-1 and -2 astronomical stations, Mini-Mega TORTORA, the Pi of the Sky observatory, TSHAO, AbAO,  RATIR, Mondy, CrAO, Maidanak and SAO RAS. See,][]{2016arXiv161203089Z,2016GCN..19588...1T}.  This burst was originally divided in three different temporal events.   \cite{2016arXiv161203089Z} stated that the spectral properties of the first two sub-bursts transition (from thermal to non-thermal radiation in a single burst) indicated the variation of the jet composition from a fireball to a Poynting-flux dominated jet.   \cite{2017arXiv170201382L} proposed that the event I could be explained by the cocoon emission surrounding the jet, the early afterglow by the superposition of the photosphere and internal shock emissions and finally, the late afterglow by the emission generated in both internal and external shocks.\\
In this paper, we use the early-afterglow external shock model in stellar wind medium and interstellar medium (ISM) to describe the multiwavelength observations during events II (henceforth called early afterglow) and III (henceforth called late afterglow) of GRB 160625B.   The paper is arranged as follows. In Section 2 we show the dynamics of external shocks that evolves adiabatically in a stellar wind-type-like medium and ISM.  In Section 3 we present the multiwavelength observations, data reduction and data analysis. In Section 4, the discussion and results on the analysis done to the multiwavelength data are presented.   Finally,  in Section 5 we give a brief summary.   
\section{Dynamics of the external shocks} 
The external shocks take place when the relativistic ejecta collide with the circumburst medium and start to be slowed down. Generally, an ongoing shock that propagates into the surrounding medium so-called forward shock and a reverse shock that propagates into the flow are formed. The afterglow phase begins when the ejecta has swept enough material so that most of the energy of the ejecta has been transferred to the circumburst medium.  We present the afterglow evolution in a stellar wind medium and ISM, and the wind-to-ISM transition.
\subsection{Afterglow evolution in the stellar wind-type-like medium}
The dynamics of a relativistic shell interacting with the surrounding medium with an inhomogeneous density  (stellar wind-like medium) has been widely discussed \citep[e.g. see,][]{2000ApJ...536..195C}.   For the adiabatic blast wave, the typical timescales (deceleration,  cooling and acceleration), the deceleration radius, the Lorentz factors,  the synchrotron spectral breaks, the maximum flux and synchrotron light curves are given in \cite{1999ApJ...520L..29C,2000ApJ...536..195C,2000ApJ...543...66P}.    Using the previous quantities, the synchrotron flux in the fast-cooling regime is proportional to {\small $\propto  t^{-\frac{3p-2}{4}}\,E^{-\frac{p}{2}} $} for {\small $E^{\rm syn}_{\rm m,f}<E^{\rm syn}<E^{\rm syn}_{\rm max,f}$} and {\small $\propto t^{-\frac{1}{4}}\,E^{-\frac{1}{2}} $} for {\small $E^{\rm syn}_{\rm c,f}<E^{\rm syn}<E^{\rm syn}_{\rm m,f}$}, where $E^{\rm syn}_{\rm c,f}$,  $E^{\rm syn}_{\rm m,f}$ and $E^{\rm syn}_{\rm max,f}$  are the synchrotron spectral breaks for the cooling, characteristic and maximum photon energy, respectively \citep[i.e.][]{2013NewAR..57..141G}.     In the slow-cooling regime, the synchrotron flux is proportional to {\small $\propto t^{-\frac{3p-1}{4}}\,E^{-\frac{p-1}{2}} $ for $E^{\rm syn}_{\rm m,f}<E^{\rm syn} <E^{\rm syn}_{\rm c,f}$} and {\small $\propto t^{-\frac{3p-2}{4}}\,E^{-\frac{p}{2}}$} for {\small $E^{\rm syn}_{\rm c,f}<E^{\rm syn}<E^{\rm syn}_{\rm max,f}$}.   Relativistic electrons accelerated in the forward shocks could scatter synchrotron photons up to energies larger than 100 MeV. The  synchrotron self-Compton (SSC) light curves with their spectral breaks are described in detail in \cite{2016ApJ...818..190F}.\\
For reverse shocks, the observables of synchrotron radiation such as  synchrotron spectral breaks, fluxes and synchrotron light curves that describe the optical flashes are derived in \cite{2000ApJ...536..195C}.   For instance, in the thick-shell regime and in the optical energy range,  synchrotron flux increases proportionally to $t^{1/2}$, reaching a maximum at  the peak time of $t_d\sim \left(\frac{\Gamma_d}{\Gamma_c}\right)^{-4}\,T_{90}$ and after the peak, it decays fast  $\propto t^{-(\beta+2)}$ dominated by the  angular time delay effect in higher latitude emissions \citep{2000ApJ...541L..51K}. Here, $\Gamma_c$ is  the critical Lorentz factor, $\Gamma_d\sim {\rm min}(\Gamma, 2\Gamma_c)$ is the bulk Lorentz factor at the shock crossing time and $T_{90}$ is the duration of the burst.  The spectral index $\beta$ corresponds to the low- and high-energy power-law indexes 1/2 and p/2, respectively.  The optical flux at the peak is derived and written explicitly in \cite{2003ApJ...597..455K}. These authors discussed the optical light curve generated in the reverse shock by the angular time delay effect produced by the high latitude emission.  For the cooling energy ($E^{\rm syn}_{\rm c,r}$) less than that characteristic energy ($E^{\rm syn}_{\rm m,r}\sim$ 1 eV), the angular time delay effect produces a peak followed by a fast decay.\\
 The observables of the SSC emission from a reverse shock such as spectral breaks, fluxes and light curves have been widely explored \citep[e. g. see,][]{2001ApJ...546L..33W,  2001ApJ...556.1010W, 2012ApJ...755...12V, 2016ApJ...818..190F}.  The  SSC light curve around the shock crossing time is derived in  \cite{2016ApJ...818..190F}.  For  $t < t_d$,  SSC  flux increases proportionally to $t^{1/2}$ and decreases ($t  > t_d$) following a fast decay $\propto t^{-(\beta+2)}$  induced by  the  angular time delay effect. Again,  the spectral index $\beta$ is 1/2 and p/2 for low and high-energy SSC power laws, respectively.  The SSC flux at the peak time $t_d\sim \left(\frac{\Gamma_d}{\Gamma_c}\right)^{-4}\,T_{90}$ is derived and explicitly calculated in \cite{2016ApJ...818..190F}.\\
The duration of the reverse-shock radiation can be calculated through $t_{\rm ang}\sim (1+z)\theta_j\,r_{d,SW}$, where $\theta_j\sim \frac{1}{\Gamma_d}\left(\frac{t_j}{t_d}\right)^{1/4}$ could be estimated, in turn, from the jet break time $t_j$ of the synchrotron flux coming from the forward shock \citep{1999ApJ...525..737R, 1999ApJ...519L..17S, 2000ApJ...541L...9K}.\\
%
%
%
\subsection{Afterglow evolution in ISM}
The dynamics of the external shocks for the ejecta expanding into a surrounding medium with homogenous density has been widely explored \citep[e.g. see,][]{1998ApJ...497L..17S}.   Using the synchrotron spectra, the evolution of synchrotron energy breaks and the maximum flux, the synchrotron light curve and spectrum in the fast-cooling regime is proportional to {\small $\propto  t^{-\frac{3p-2}{4}}\,E^{-\frac{p}{2}} $} for {\small $E^{\rm syn}_{\rm m,f}<E^{\rm syn}<E^{\rm syn}_{\rm max,f}$} and {\small $\propto t^{-\frac{1}{4}} \,E^{-\frac{1}{2}}$} for {\small $E^{\rm syn}_{\rm c,f}<E^{\rm syn}<E^{\rm syn}_{\rm m,f}$} \cite[i.e.][]{2010ApJ...722..235V, 2013NewAR..57..141G}.  In the slow-cooling regime, the synchrotron light curve and spectrum is proportional to {\small $\propto t^{-\frac{3p-3}{4}}\,E^{-\frac{p-1}{2}}$ for $E^{\rm syn}_{\rm m,f}<E^{\rm syn} <E^{\rm syn}_{\rm c,f}$} and {\small $\propto t^{-\frac{3p-2}{4}}\,E^{-\frac{p}{2}}$} for {\small $E^{\rm syn}_{\rm c,f}<E^{\rm syn}<E^{\rm syn}_{\rm max,f}$}, where the proportionality constants of these spectra are explicitly written in e.g., \citet{2016ApJ...831...22F}. \\ 
\\
The achromatic break in the optical and X-ray bands observed in the late afterglow is related to the time when the jet   slows down and spreads laterally \citep{1999ApJ...519L..17S}. For this case, assuming  the synchrotron emission from the same power-law electron distribution and also that the jet break ($\Gamma\sim\theta_j^{-1}$) takes place at time  $t_{\rm j}\propto (1+z)\,n^{-1/3}\,E^{1/3}\,\theta_j^{8/3}$,  the synchrotron flux for slow-cooling regime becomes: $\propto t^{-p}\,E^{-\frac{p}{2}}$ for $E^{\rm syn}_{\rm c,f}<E^{\rm syn}$, $\propto t^{-p}\,E^{-\frac{p-1}{2}}$ for  $E^{\rm syn}_{\rm m,f}<E^{\rm syn}<E^{\rm syn}_{\rm c,f}$, and $\propto t^{-1/3}\,E^{\frac{1}{3}}$ for $E^{\rm syn}<E^{\rm syn}_{\rm m,f}$ \citep{1999ApJ...519L..17S}. \\ 
%
%
%
%
\vspace{1cm}
\subsection{The Wind-to-ISM Transition}
%
%
%
LGRBs are thought  to be associated with the core collapse of massive stars, suggesting that the medium surrounding the progenitor is modified by the stellar wind. In the case of a  Wolf-Rayet,  for a mass-loss rate $\dot{M}\simeq 10^{-6}\,{\rm M_\odot\,yr^{-1}}$ with a wind velocity constant of $\rm v_W\simeq 10^8\, {\rm cm\, s^{-1}}$, the density of the stellar wind-type-like medium is given by  $\rho(r)=A\, r^{-2}$, where  $A=\frac{\dot{M}}{4\pi \rm v_W}= A_{\star}\,\, (5\times 10^{11})\, {\rm g\, cm^{-1}}$ with $A_{\star}$ a parameter of stellar wind density \citep{2000A&A...362..295V, 2005A&A...442..587V, 2004ApJ...606..369C, 1998MNRAS.298...87D, 2003ApJ...591L..21D}.  The dynamics of the wind-to-ISM transition phase was originally introduced by \cite{1977ApJ...218..377W} and \cite{1975ApJ...200L.107C}.  Authors showed that this phase was made up of  four-region structure which are (1) the unshocked stellar wind with density $\rho(r)$, (2) a quasi-isobaric zone consisting of the stellar wind mixed with a small fraction of interstellar gas, (3) a dense-thin shell formed by most of ISM and (4) the unshocked ambient ISM (see Figure 1 in \citealp{2006ApJ...643.1036P}).\\
%
%
Taking into consideration an adiabatic expansion, two strong shocks are formed, the outer and inner shocks. The outer termination (forward) shock radius can be estimated as 
{\small
\bary
R_{FS, W}&=&\left(\frac{125}{308\,\pi} \right)^\frac15\left(\frac{\dot{M}\,\rm v^2_W t^3_\star }{\rm n} \right)^\frac15\cr
&=&  1.2\times 10^{19}\,  {\rm cm}\,\,\dot{M}^\frac15_{-6}\,\rm v_{W,8}^\frac25\,n^{-\frac15}_0\,t_{\star,5}^{-\frac35}\,,
\eary
}
where $t_\star$ is the lifetime of the Wolf-Rayet phase of the star and the homogeneous density has been written as   $\rm {n=n_0}\,\,{\rm cm^{-3}}$.\\
The inner (reverse) shock radius for which the wind-to-ISM transition takes place \citep[$R_0$;][]{2006ApJ...643.1036P} is calculated equaling the pressures in zones (2) and (3)
{\small
\bary
P_{\rm (2)}=P_{\rm (3)}&=&\frac{7}{25} \left(\frac{125}{308\,\pi} \right)^\frac25 \left(\frac{\dot{M}\,\rm v^2_W }{\rm n\,t^2_\star} \right)^\frac25\cr
&=&  1.4\times 10^{-11}\,{\rm dynes\, cm^{-2}}\,\,\dot{M}^\frac25_{-6}\,\rm v_{W,8}^\frac45\,n^{-\frac{3}{5}}_0\,t_{\star,5}^{\frac45}\,,
\eary
}
with the strong conditions at the shock  \citep[e.g. see, ][]{2006ApJ...643.1036P, 1996ApJ...469..171G, 2002ApJ...565L..87D}. In this case the inner shock and the  wind-to-ISM transition radius can be written as

{\small
\bary
R_0\equiv R_{RS,W}&=&\left(\frac{3\,\dot{M}\,v_W }{16\pi P_{\rm (2)}} \right)\cr
&=&  5.1\times10^{18} {\rm cm} \,\,\dot{M}^\frac{3}{10}_{-6}\,\rm v_{W,8}^\frac{1}{10}\,n^{-\frac{3}{10}}_0\,t_{\star,5}^{\frac25}\,.
\eary
}
The density of stellar wind  at $r=R_0$ is given by
{\small
\bary
\rho(R_0)&=&\frac{\dot{M}}{4\pi R^2_0\, \rm v_w}\cr
&=& 1.8\times 10^{-27}\, {\rm g\,cm^{-3}}\,\,  R^{-2}_0\,\dot{M}_{-6}\, \rm v_{W,8}^{-1}\,,
\eary
}
which corresponds to a number density of particles $1.1\times 10^{-3}\, {\rm cm^{-3}}$.
\section{GRB160625B: Multiwavelength observations, data reduction and data analysis}
\subsection{Multiwavelength observations and data reduction}
At 22:40:16.28 UT, 2016 June 25, Fermi-GBM triggered and located GRB 160625B \citep{2016GCN..19581...1B}.  Later, at 22:43:24.82 UT, Fermi-LAT  triggered on  a luminous pulse of the ongoing burst.   More than 300 photons were detected above 100 MeV in the direction of this burst and  the highest-energy photon detected  was 15 GeV  observed at 345 s after the GBM trigger \citep{2016GCN..19586...1D}.    XRT on board the Swift satellite followed up this burst for $\sim$ 1.1 ks \citep{2016GCN..19585...1M}.  Surprisingly, at 22:51:16.03 GBM again triggered on this burst.   Several optical observations were performed with  the Pi of the Sky observatory,  Mini-Mega TORTOLA,  TSHAO, AbAO,  Mondy, CrAO, Maidanak and SAO RAS \citep[See; ][]{2016arXiv161203089Z} and RATIR instrument \citep[{\rm riZYJH} bands;][]{2016GCN..19588...1T}.  This burst also triggered Konus-Wind  at 22:40:19.875 UT.   Assuming the redshift z=1.406 \citep{2016GCN..19600...1X},   Konus-Wind measured the highest isotropic energy ever detected of $\sim 4\times 10^{54}$ erg \citep{2016GCN..19604...1S}.\\
Fermi-LAT data in the energy range of 100 MeV - 300 GeV  was reduced using the public database at the Fermi website\footnote{http://fermi.gsfc.nasa.gov/ssc/data}.  The lightcurve was obtained using the  ScienceTools-v9r27p1 package and the P7TRANSIENT V6 response function.   The Swift-XRT data used in this work are publicly available at the official Swift web site\footnote{http://swift.gsfc.nasa.gov/cgi-bin/sdc/ql?}.  The optical fluxes and their associated errors  used in this work were calculated using the magnitudes reported by \cite{2016arXiv161203089Z} with the standard conversion for AB magnitudes shown in \cite{1996AJ....111.1748F}.
\subsection{Multiwavelength data analysis}
\begin{figure}
\epsscale{.80}
\plotone{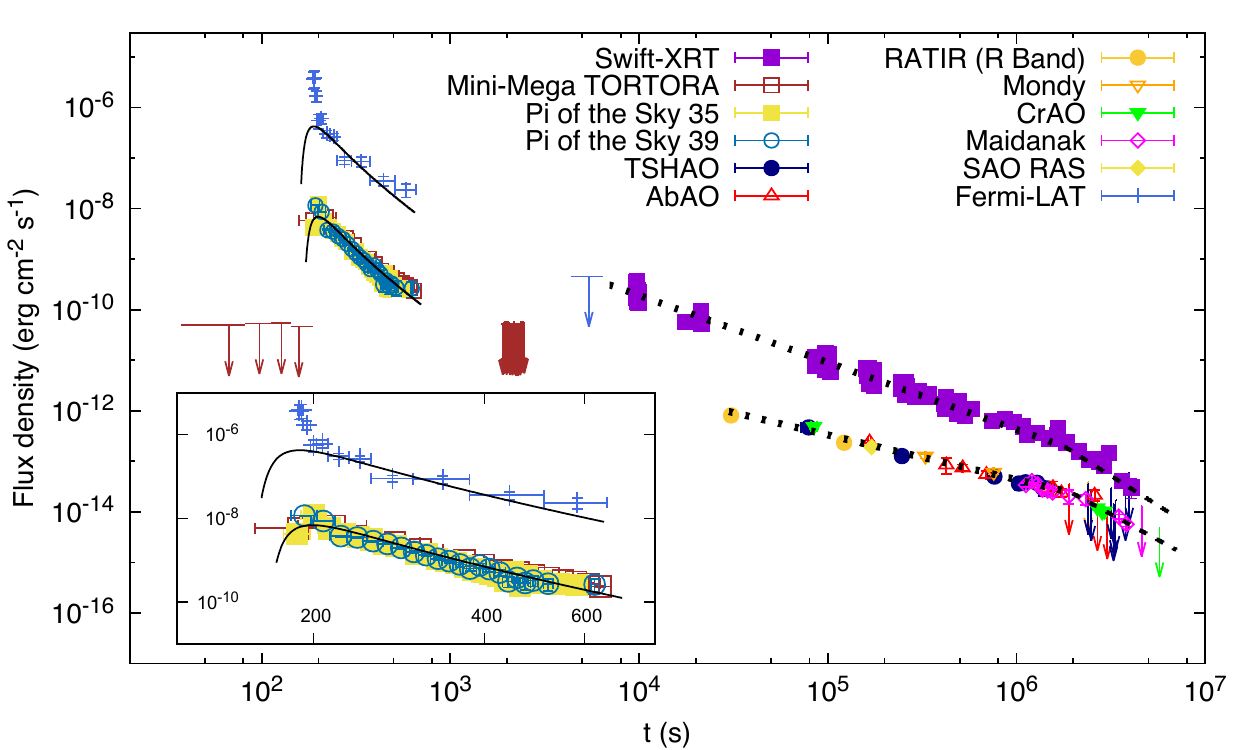}
\caption{Light curves and fits of the multiwavelength (LAT, XRT and UVOT) observation of GRB 160625B with our external shock model.  We use the LC of  RS  in the thick-shell regime to describe the bright LAT-peak flux (continuous line), the LC of FS to explain  the temporally extended LAT, X-ray and optical emissions  before the break time at $t_{br}\sim\,1.5\times 10^3\, {\rm s}$  (dotted lines) and the LC  after the jet break time (dashed lines).    Fermi-LAT data  were reduced using the public database at the Fermi website, the Swift-XRT data were obtained using the  publicly available database at the official Swift web site, optical data (Mini-Mega TORTORA, Pi of the Sky 35 and 39, TSHAO, AbAO,  Mondy, CrAO, Maidanak and SAO RAS) were collected from \cite{2016arXiv161203089Z} and RATIR R-band data were obtained from \cite{2016GCN..19588...1T}.  The initial time of this plot is the onset of the small precursor event detected by the Fermi-GBM}.
\label{fit_afterglow}
\end{figure}

\begin{figure}
\epsscale{1.2}
\plotone{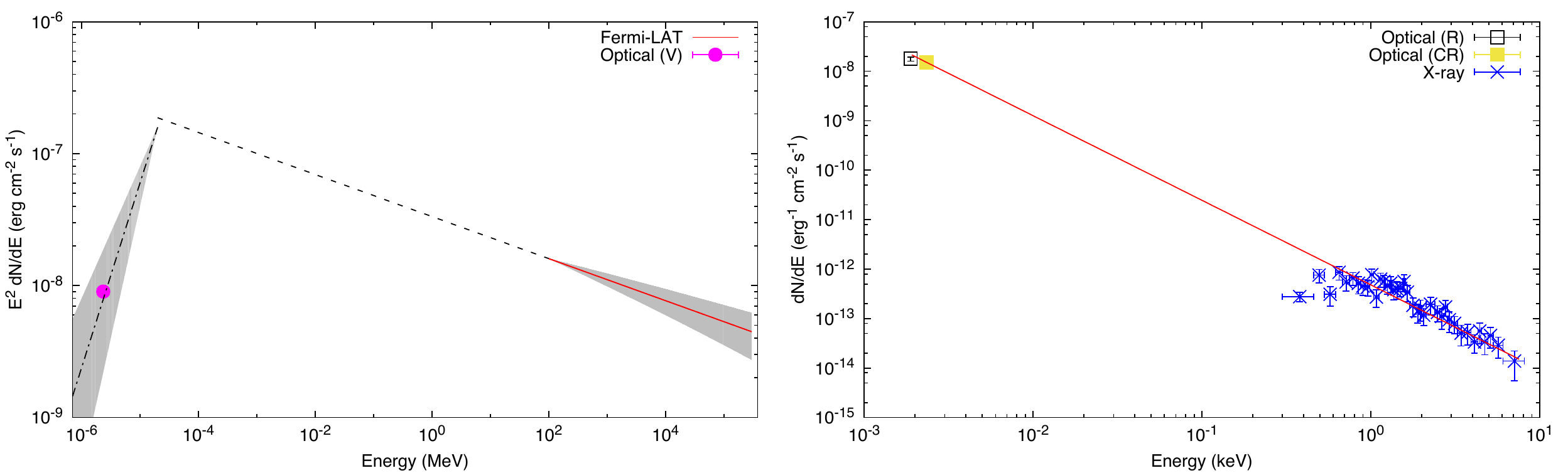}
\caption{Left panel:  SED of the Fermi-LAT and optical observations of the early afterglow.  Right panel: SED of the X-ray and optical observations of the late afterglow.  Optical R-,  CR- and V-band data were average over each band.}
\label{fit_spectrum}
\end{figure}

The Chi-square $\chi^2$ minimization using  the ROOT software package \citep{1997NIMPA.389...81B} was done to fit the multiwavelength observations presented in the early and late afterglow.  The values observed of decay slopes with their chi squares  ($\chi^2$ / n.d.f.\footnote{Number of degrees of freedom}) are reported in Table 1.\\
%
%
%
Due to connection between prompt emission and the early afterglow,  we model the fast decay of the optical and LAT fluxes using the function
\be
F(t)= A \left(\frac{t-t_0}{t_0}\right)^{-\alpha}\,e^{-\frac{\tau}{t-t_0}}\,,
\ee
where $t_0$ is the  starting time,  $A$ is the amplitude, $\tau$ is the timescale of the flux rise and $\alpha$ is the temporal decay index  \citep{2006Natur.442..172V}.   A blow-up of the optical and LAT light curves together with the modeling function before $\sim$ 700 sec  is shown in Figure \ref{fit_afterglow}.   The best-fit values obtained of the optical (LAT) flux were $t_0=153.3\pm22.1 (142.4\pm9.8)$ s,  $\tau=101.2\pm9.3(95\pm3.4)$ s and  $\alpha=2.51\pm0.81(2.46\pm0.75) $.  The values of starting times suggest that both (LAT and optical) afterglow emission started simultaneously around  $\sim$ 150 s.   If a wrong $t_0$ is chosen to the precursor time, then some unreasonable results are obtained. However, if $t_0$ is chosen at the main burst, then  the reasonable results as presented in the paper are obtained. This is understandable since the precursor is energetically insignificant. The blast wave dynamics is mostly defined by the main burst.   On the other hand, the Fermi-LAT spectrum was plotted and modeled with a power law function (see Figure \ref{fit_spectrum}; left panel).   The best-fit value found of the LAT spectral index is $\Gamma_{\rm LAT}=\beta_{\rm LAT}+1=$2.15$\pm$ 0.05 and the observational LAT flux as function of time and energy is  $F_\nu\propto t^{-2.46\pm0.75}\, E^{-1.15\pm0.05}$.  Taking into consideration that during the early afterglow, this burst was only detected in the optical V-band, an extended dotted-dashed line over the only optical data was drawn.   The value of the slope of $1.45\pm0.05$ for this line was chosen in accordance with the closure relation in our model. Deviations from those relations have been extensively analyzed 
in \cite{2014ApJ...780...82U}.    In this case, optical flux varies as  $F_\nu\propto t^{-2.51\pm0.81}\, E^{-0.45\pm0.05}$. \\
%
%
%
It is worth noting that the optical band is typically in the regime $E^{\rm syn}_{\rm m,r}(t_d)<E^{\rm syn}_{\rm opt}< E^{\rm syn}_{\rm c,r}(t_d)$ for ISM \citep{2005ApJ...628..315Z, 2003ApJ...582L..75K} and $E^{\rm syn}_{\rm c,r}(t_d)<E^{\rm syn}_{\rm opt}< E^{\rm syn}_{\rm m,r}(t_d)$ for stellar wind-type like medium \citep{2003ApJ...597..455K}.  After the peak,  the flux at an energy above $E^{\rm syn}_{\rm c,r}$ disappears at $t_d$ because no electron is shocked anymore.  For ISM,  the cooling break energy is larger than the optical band ($E^{\rm syn} < E^{\rm syn}_{\rm c,r}$) and the optical flux  decays $\propto t^{-(73p+21)/96}\simeq t^{-2}$ \citep{2000ApJ...545..807K}. For stellar wind medium,  the cooling break energy is smaller than the optical band  ($E^{\rm syn}_{\rm c,r}< E^{\rm syn}$) and the optical flux  decays  $\propto t^{-(\beta+2)}$ when the angular time delay effect produced by high latitude emission is considered  \citep{2003ApJ...597..455K, 2000ApJ...541L..51K}.  Taking into consideration the value of spectral index $\beta_{\rm LAT}=1.15\pm 0.05$,  the LAT and optical fluxes are consistent with synchrotron and SSC emission in the fast-cooling regime for p=2.4 when the outflow is decelerated by the stellar wind medium.    In order to find a correlation between GeV $\gamma$-ray and optical fluxes, the Pearson's correlation coefficients with the {\rm  p}-values are calculated.   Considering a maximum allowed time difference between data of  $\Delta t\lesssim$ 2, 5 and 10 s, the Pearson's correlation coefficients are 0.93, 0.91 and 0.92  and the p values are $2.2\times 10^{-2}$ (i.e. the probability of being by chance is 1.2\% ), $6.6\times 10^{-4}$ and $1.1\times 10^{-8}$, respectively.    The values of these coefficients obtained during the period in which both GeV $\gamma$-ray and optical fluxes began to decline, reveal that GeV $\gamma$-ray and optical fluxes are strongly correlated and also that this correlation is not due to random chance. 
The observational (spectral an temporal) and  theoretical values of the decay slopes (see Table 1) and the strong correlation between both fluxes are consistent with  the theoretical values  of synchrotron and SSC radiation from the reverse shock evolution in the stellar wind-type-like medium. This evolution corresponds to a thick-shell regime affected by the angular time delay effect \citep[see, ][]{2003ApJ...597..455K, 2000ApJ...541L..51K}.\\
%
%
%
%
%
%
\\ 
\\
During the late afterglow from $\sim 8\times 10^3$  to $6\times 10^6$ s, X-rays and optical lightcurves were observed with a break at $t_j\sim1.6\times 10^{6}$ s. The slopes of the X-ray  and optical fluxes before the break are $\alpha_{\rm X,bb}=1.327\pm 0.521$ and $\alpha_{\rm opt,bb}=0.921\pm 0.163$, and after the breaks are $\alpha_{\rm X,ab}=2.348\pm 0.860$ and $\alpha_{\rm opt,ab}=2.036\pm 0.521$, respectively.  In addition, the spectrum energy distribution (SED) of the optical and X-ray data was modelled with a power law function (see Figure \ref{fit_spectrum}; right panel) and the best-fit value of  $\Gamma_{\rm X,opt}=\beta_{\rm X,opt}+1=$1.71$\pm$ 0.12 was obtained.  Therefore, the flux varies as  $F_\nu\propto t^{-1.327\pm 0.521}\, E^{-0.71\pm0.12}$ and  $F_\nu\propto t^{-0.921\pm0.163}\, E^{-0.71\pm0.12}$ for X-ray and optical wavelengths, respectively.  These results indicate that  the slopes observed for the X-ray and optical fluxes before the breaks are consistent with the forward-shock synchrotron emission in the slow-cooling regime ($E^{\rm syn}_{\rm m,f}<E^{\rm syn}<E^{\rm syn}_{\rm c,f}$) for a power-law index of p=2.4 when outflow is decelerated by the ISM. After the breaks,post jet-break fluxes are proportional to $F_\nu\propto t^{-2.348\pm 0.860}\, E^{-0.71\pm0.12}$ and  $F_\nu\propto t^{-2.036\pm 0.521}\, E^{-0.71\pm0.12}$ for X-ray and optical wavelengths, respectively, which are consistent with synchrotron radiation in the slow-cooling regime for p=2.4. The observational and theoretical values are reported in Table 1.\\
%
%
%
%
%
%
In general, using the reverse- and forward-shock light curves it can be seen that the early afterglow is consistent with the afterglow evolution in the wind medium and the late afterglow is consistent with the afterglow evolution in ISM.  Table 1 shows that both values of slope decays (observational and theoretical) are in agreement.\\
%
\\
%
Taking into account the starting time found of the LAT and optical afterglow $t_0\approx$ 150 s, the values of the bulk Lorentz factor $\Gamma$ and the parameter $A_\star$ are constrained through the deceleration time  in wind-type  like medium  $t_{\rm d, SW}(\Gamma, A_\star)\approx$ 150 s.  Taking into consideration that the early and late afterglows are consistent with radiation emitted when ejecta are decelerated in the  stellar wind density and ISM, respectively,  the wind-to-ISM transition must have taken place between $\sim$ 700 s and $\sim 10^4$ s (see Fig. \ref{fit_afterglow}).\\
%
%
 %
%
%
By using the values of the isotropic radiated energy  $\sim 4\times 10^{54}$ erg \citep{2016GCN..19604...1S} with an efficiency $\eta\approx$ 0.2 which corresponds to a kinetic energy of $2\times 10^{55}$ erg, the redshift z=1.406 \citep{2016GCN..19600...1X}, the spectral index of electron distribution $p=2.4$ and  the duration of the burst $T_{90}\simeq$188 s,  the fit in the early phase of the LAT and optical early data were done with synchrotron and SSC model in the stellar wind-type-like medium for a relativistic electron population radiating photons at 180 s with energies of 100 MeV and 2 eV, respectively. The late phase of the X-ray and optical data were modelled with synchrotron  emission for the same relativistic electron population radiating photons at $5\times 10^{4}$ s with energies of 5 keV and 2 eV, respectively.  For  t $\geq t_{\rm j,br}\simeq 1.6\times 10^6$ s, the post jet-break synchrotron light curves  in the slow cooling regime  for X-ray and optical fluxes are used.\\ 
\\
\section{Results and Discussion}
The multiwavelength data of the early afterglow (GeV and optical bands) and the late afterglow (X-rays and optical bands) are shown in Figure \ref{fit_afterglow}.  In addition, we show the fit of the early afterglow  using the afterglow evolution in the stellar wind-type-like medium and the late afterglow  using the afterglow evolution in ISM.     The values of the microphysical parameters and densities found with the fit of the multiwavelength observations and the wind-to-ISM transition are reported in Table 2.  Using the values of the parameters in Table 2,  we can infer the following:
\\
\begin{enumerate}
\item Using the value found of the magnetic microphysical parameter after describing the early afterglow, the magnetization parameter becomes $\sigma\simeq$ 0.4.   This value means that ejecta is moderately magnetized and therefore, a successful reverse shock is expected. Otherwise, particle acceleration in the reverse shock is inefficient and the reverse shock would have been suppressed  (for $\sigma\gg$1).  In addition, for  $\sigma\gg$1 the bright optical and LAT peaks would not have been detected \citep{2005ApJ...628..315Z, 2004A&A...424..477F}.  Several authors have pointed out that Poynting flux-dominated models with arbitrary magnetization could give account of the high-energy emission observed in the brightest LAT-detected bursts \citep{2014NatPh..10..351U,2011ApJ...726...90Z}.    This value  indicates that the ejecta must also have dissipated a significant amount of Poynting flux during the prompt emission  phase, being  the internal collision-induced magnetic reconnection and turbulence (ICMART) event  the most favorable process to explain this pattern \citep{2011ApJ...726...90Z}.  This result agrees with the model proposed by \citet{2016arXiv161203089Z} after analyzing the spectral properties exhibited in GRB 160625B. They suggested that the thermal and non-thermal emission coming from the events I and II could be explained through the transition from a fireball to Poynting flux-dominated jet.
\item The values of wind ($A_\star=0.2$) and ISM (n=10 cm$^{-3}$)  parameter densities found for the early and late afterglow, respectively, lie in the range of typical ones reported for highly energetic burst \citep{2013ApJ...763...71A, 2014ApJ...781...37P, 2014Sci...343...38V, 2012ApJ...751...33F, 2008Natur.455..183R}.  The values of circumburst medium  n=10 cm$^{-3}$ and the distance z=1.406 associated with this burst support the idea that the host could be  a dwarf-irregular galaxy which has typical size of L $\sim$ 0.1 kpc and column density of $N_H\simeq 3\times 10^{21}\,{\rm cm^{-2}}$ \citep{1998ApJ...507L..25B, 2001ApJ...554..678B}.
\item Using the values of the parameters ($A_\star=0.2$) and (n = 10 cm$^{-3}$),  we derived the values of the wind-to-ISM transition for the deceleration time in the stellar wind and the transition radius which are  $t_{\rm d, W}=7.8\times 10^{3}$ s  and  $R_{\rm 0}\simeq2.6\times 10^{18}$ cm, respectively.  The value of initial bulk Lorentz factor derived in this afterglow model corresponds to $\Gamma$=500 similar to the LAT-detected bursts.   By studying the spectral features of the LAT-detected bursts, \cite{2012ApJ...755...12V} used a magnetically dominated ejecta model to describe the high-energy emission present in these energetic bursts.  They showed that the inverse Compton scattering coming from the forward and reverse shocks give a significant contribution in the LAT emission, provided that the bulk Lorentz factor were in the range of 300- 600. In addition, other powerful bursts such as GRBs 110731A and 130427A were modelled using synchrotron and SSC emission from the external shock model for bulk Lorentz factors of 520 and 550, respectively. Considering that GRB 160625B is among the five most powerful bursts, it is expected that the value of the bulk Lorentz factor is in the range of the brightest LAT-detected bursts, as found in this work.
\end{enumerate}
Table 3 shows the timescales, bulk Lorentz factors,  synchrotron and SSC spectral breaks among others.  These values were computed based on the values reported in Table 2 and the dynamics of a relativistic shell interacting with an stellar wind medium \citep{2000ApJ...536..195C} and ISM \citep{1998ApJ...497L..17S} for the early and late afterglows, respectively.  In accordance with the quantities reported in Table 3,  the following results are found:\\
\\
\begin{enumerate}
\item By comparing the synchrotron self-absorption energy ($E^{\rm syn}_{\rm a,r}$) with the characteristic ($E^{\rm syn}_{\rm m,r}$) and cooling ($E^{\rm syn}_{\rm c,r}$) energies obtained the early afterglow, it can be noted that synchrotron spectrum in the reverse shock lies in the weak self-absorption regime.  Therefore, a thermal component generated by synchrotron radiation is not expected at $\sim$ 150 - 200 s.
\item The break observed in X-ray and optical light curves  at $t_j\simeq 1.6\times 10^6$ s is attributed to a jet break, leading to a jet opening angle of $\theta_j\simeq 8.3^\circ$ \citep{1999ApJ...519L..17S}.   The value of  bulk Lorentz factor at the jet-break time corresponds to $\Gamma_{\rm j, br}$ = 6.9. The beaming corrected gamma-ray energy is then $3\times10^{52}$ erg which makes it part of the hyper energetic GRBs \citep{2011ApJ...732...29C}.
\item The maximum energy of synchrotron photons radiated in the stellar wind afterglow (the second event) is 7.69 GeV at 350 s.  Then, the highest-energy photon of 15 GeV detected at 354 s after the GBM trigger is not consistent with the maximum 
synchrotron energy from an adiabatic forward shock  propagating into the stellar wind of the star. Therefore,  the most energetic photon could be explained by the inverse Compton scattering from the forward shock which has a characteristic break energy of 7.3 TeV.
\item During the early afterglow, a temporal correlation was found between the GeV $\gamma$-ray and optical bands. It suggests that the GeV $\gamma$-ray and optical fluxes were generated co-spatially by the same electron population.   During the prompt and afterglow phases,  correlations among distinct bands have been searched in order to explore the origin of the LAT-detected emission.  For instance,   an optical flash observed in the extremely brightest GRB 130427A  correlated with  the LAT-detected emission, indicating that both emissions originated in the early afterglow.  A very similar pattern is found in GRB 160625B which displayed a bright optical flash in temporal correlation with the LAT emission.  It suggests that the LAT emission could have been generated in the early afterglow.
\item  The ratio of the magnetic fields in the forward- and reverse-shock regions found is  $B_r/B_f\simeq 620$.  The magnetic field in the reverse shock region is stronger than in the forward shock  as found in the brightest LAT-detected burst (GRB 090510,  GRB 110721A, GRB 110731A, GRB 130427A and others) the ejecta is magnetised. As follows we estimate the synchrotron flux contribution from the reverse and forward shocks. The forward-shock synchrotron quantities at $\sim$ 200 s inferred from the  later times are:  $E^{\rm syn}_{\rm m,f}=1.4\, {\rm keV}$, $E^{\rm syn}_{\rm c,f}=0.9\, {\rm eV}$ and $F^{\rm syn}_{\rm max,f}=31.9\,{\rm mJy}$,  at {\small $E^{\rm syn}=$} 2 eV,  flux is in the energy range of $E^{\rm syn}_{\rm c,f} < E^{\rm syn} <E^{\rm syn}_{\rm m,f}$, and then it is given by  {\small $F_{\rm \nu,f}=F^{\rm syn}_{\rm max,f} \left(E^{\rm syn}/E^{\rm syn}_{\rm c,f}\right)^{-1/2}$}  \citep{1998ApJ...497L..17S}.  From the  reverse-shock synchrotron quantities reported in Table 3, the reverse-shock synchrotron flux at {\small $E^{\rm syn}=$} 2 eV lies in the energy range of $ E^{\rm syn}_{\rm m, r} < E^{\rm syn}$.   Therefore,  it can be written as {\small $F_{\rm \nu,r}=F^{\rm syn}_{\rm max, r} \left(E^{\rm syn}_{\rm m, r}/E^{\rm syn}_{\rm c, r}\right)^{-1/2}  \left(E^{\rm syn}/E^{\rm syn}_{\rm m, r}\right)^{-p/2}$} \citep{1998ApJ...497L..17S}.   The synchrotron fluxes at forward and reverse shocks are $F_{\rm \nu,f}$ = 21.4 mJy and $F_{\rm \nu,r}$ = 11.2$\times 10^2$ mJy, respectively.   These values in the fluxes indicate that synchrotron emission from the reverse shock is dominant over that radiation originated at the forward shock. The previous results together with the fact that  the polarization percentage from the forward shocked circumburst medium is expected to be very low \citep[see e.g.;][]{1999A&A...348L...1C, 2003Natur.426..157G} suggest that the optical flux is expected with some degree of polarization.\\   
\item GRB 160625B is one of the most energetic burst, suggesting a large amount of target photons for photo-hadronic interactions and then, making it a potential candidate for neutrino detection.  However,  no high-energy neutrinos in spatial and temporal coincidences were reported by the IceCube neutrino telescope around this burst.    A similar powerful burst GRB 130427A with energy of $\sim\,2\times 10^{54}$ erg  was detected by several  satellites and ground-based telescopes \citep{2014Sci...343...48M, 2014Sci...343...42A, 2014Sci...343...38V} and although  searches for TeV - PeV neutrinos were performed,  no excess were found above background.  \cite{2013ApJ...772L...4G} stated that the neutrino non-detection could constrain the values of the bulk Lorentz factor, emitting radius and the energy fraction converted into cosmic rays $\epsilon_p$.  They found that almost independently of the bulk Lorentz factor, the energy fraction between electrons and cosmic rays lies in the range $\epsilon_p\lesssim  \,\epsilon_e$.  Although a robust analysis could be required,  a simple proof can be done for GRB 160625B following a similar procedure.  Form our results obtained in early afterglow can be seen that the energy fraction given to accelerate electrons and amplified the magnetic field at the end of the prompt phase is $\epsilon_e=$ 0.5 and  $\epsilon_{B,r}=$ 0.4, respectively.   Taking into consideration the energy conservation condition $\epsilon_{B,r}+ \epsilon_{e,r}+\epsilon_p\lesssim1$, then  the energy fraction converted into cosmic rays would be limited by $\epsilon_p\lesssim  \frac12\,\epsilon_e$. This result is very similar to that found by  \cite{2013ApJ...772L...4G} for GRB 130427A and  might explain the lack of high-energy neutrinos around GRB 160625B.
\item  \citet{2010ApJ...720.1513K} studied the optical photometry data in a total of 42 GRB afterglows. They found that 10\% of the afterglows presented optical peaks followed by a fast decay which are usually associated with a reverse shock flash. Several authors have claimed that this kind of afterglow, as observed in GRB080319B \citep{2008Natur.455..183R}, GRB130427A \citep{2014Sci...343...42A, 2014Sci...343...38V}, GRB050904 \citep{2007AJ....133.1187K}, GRB120711A \citep{2014A&A...567A..84M}, and GRB990123 \citep{1999Natur.398..400A}, among others, are only present in the most luminous bursts. Given that GRB 160625B has been one of the most powerful bursts detected which exhibited an optical flash with a fast decay, this burst seems to confirm this statement and belong in the same category.
 \end{enumerate}
\vspace{0.8cm}
\section{Conclusions}
We have described the non-thermal multiwavelength observations of GRB 160625B collected with Fermi-LAT, Swift-XRT and several optical ground observatories.  The multiwavelength observations of the early afterglow are consistent with the afterglow evolution in a stellar wind medium.    The optical spectral index is consistent with the synchrotron radiation while GeV $\gamma$-ray flux  with SSC emission dominated by the high latitude emission.  In this event, a strong correlation between GeV $\gamma$-ray and optical fluxes was found.  On the other hand,  the multiwavelength observations of the late afterglow are consistent with the afterglow evolution in ISM instead of the stellar wind profile.  The X-ray and optical spectral indices in this event are consistent with synchrotron radiation from the adiabatic forward shock. The X-ray and optical flux decay indices after the break time of $\sim 1.6 \times 10^6$ s are softer than forward-shock synchrotron emission, being more consistent with the evolution of the jet after reaching a jet break. Using the observed jet break time  of $\sim 1.6\times 10^6$ s in X-ray and optical light curves, the opening angle of the jet and the bulk Lorentz factor at the jet break found are  8.3$^\circ$  and 6.9, respectively.\\
Optical and GeV $\gamma$-ray fluxes of the early afterglow were modeled with synchrotron and SSC emission from reverse shocks when the ultra-relativistic electrons are accelerated in the reverse shock evolving in the thick-shell regime.   Optical and X-ray fluxes of late afterglow were fitted with synchrotron radiation from the adiabatic forward shocks.   The inverse Compton scattering process from forward shock must be included in this afterglow model  in order to explain the highest-energy photon of 15 GeV detected at 354 s after the GBM trigger.  The values found of the wind density and ISM parameters are $A_\star=0.2$ and ${\rm n= 10\,cm^{-3}}$. The value of ISM parameter found of this burst supports the idea that the host could be  a dwarf-irregular galaxy. The values obtained in wind-to-ISM transition for the deceleration time in the stellar wind and transition radius are  $t_{\rm d, W}=7.8\times 10^3$ s  and  $R_{\rm 0}\simeq2.6\times 10^{18}$ cm, respectively.\\
The value of the magnetization parameter $\sigma\simeq 0.4$ found after modelling the GeV $\gamma$-ray and optical fluxes in the early afterglow indicates that  the Poynting flux-dominated jet models with arbitrary magnetization could give account about the spectral properties exhibited in GRB 160625B.  Taking into consideration that the ejecta must  be magnetized  and  the synchrotron emission from the reverse shock is stronger than the radiation originated in the forward shock,  then  optical polarization is expected from the reverse-shock region.\\
The value found of the initial bulk Lorentz factor $\Gamma\simeq500$, and the bright optical flash with a fast decay reported in this burst indicates that GRB 160625B shares similarities with the most luminous LAT and pre-LAT era events, consistent with GRB160625B being one of the most extreme GRBs regarding energy output.\\
\acknowledgments

We thank the anonymous referee for valuable suggestions that helped improve our manuscript.   We thank Simone Dichiara, Fabio De Colle and Alexander A. Kann for useful discussions.  NF  acknowledges  financial  support  from UNAM-DGAPA-PAPIIT  through  grant  IA102917  and WHL through grant 100317. PV thanks Fermi grant NNM11AA01A and partial support from OTKA NN 111016 grant. 
%
%

%
\clearpage
\clearpage
\begin{center}\renewcommand{\arraystretch}{0.7}\addtolength{\tabcolsep}{-4pt}
\begin{center}
\scriptsize{\textbf{Table 1. Fitted values of the multiwavelength data.  Values in round parenthesis are the chi-square minimization ($\chi^2$ / n.d.f.)}} \\
\end{center}
\begin{tabular}{ l c c c c}
\hline
\hline
        		                           &										&\hspace{0.5cm} 	Early afterglow			 &\hspace{0.5cm} 	Late afterglow	&		 \\ 
		                           &										&\hspace{0.5cm}  \scriptsize{Observation}\hspace{1.5cm}  \scriptsize{Theory} \hspace{0.5cm}&\hspace{0.5cm}  \scriptsize{Observation}\hspace{1cm} \scriptsize{Theory}  &       \\
		                           
\hline \hline
\scriptsize{\bf{GeV flux}} 	& 			 				                         &	  		 	         &	   & 			 \\ 
\hline \hline  
\scriptsize{Decay slope}	&  \scriptsize{$\alpha_{LAT}$\hspace{0.5cm}}	 	&		\scriptsize{$2.46\pm0.75$  (44.77/10)\hspace{0.5cm}   3.20      \hspace{0.5cm}}&		\scriptsize{$ - $   \hspace{1cm}}	\\
\scriptsize{Spectral slope}	&  \scriptsize{$\beta_{LAT}$\hspace{0.5cm}}	 	&		\scriptsize{$1.15\pm0.05$  (35.6/31)\hspace{0.6cm}    1.20      \hspace{0.7cm}}&		\scriptsize{$ - $   \hspace{1cm}}	\\

\\
\hline \hline
\scriptsize{\bf{X-ray flux}}		         & 		                                                                                  \\
\hline \hline
\scriptsize{Decay slope (before break)} &\scriptsize{$\alpha_{\rm X,bb}$\hspace{0.5cm}}			 &			\scriptsize{$-$\hspace{0.5cm}}	&\scriptsize{$1.327\pm0.521$ (156.5/112) \hspace{0.5cm}   1.05 } \\
\scriptsize{Decay slope (after break)}    &\scriptsize{$\alpha_{\rm X,ab}$\hspace{0.5cm}}			 &			\scriptsize{$-$\hspace{0.5cm}} & \scriptsize{$2.348\pm0.860$ (4.869/6) \hspace{0.7cm}   2.40}	\\
\scriptsize{Break time (s)}&\scriptsize{$t_{\rm X, br}$ \hspace{0.5cm}}			 &		\scriptsize{ $-$\hspace{0.5cm}}  & \scriptsize{$1.64 \times 10^6$} 	\\
\scriptsize{Spectral slope }    &\scriptsize{$\beta_{\rm X}$\hspace{0.5cm}}			 &			\scriptsize{$-$\hspace{0.6cm}} & \scriptsize{$0.71\pm0.12$ (156.5/112) \hspace{0.7cm}   0.70}	\\
\\
\hline\hline
\scriptsize{\bf{Optical flux}}  		         &                                                                                  \\
\hline\hline
\scriptsize{Early decay slope } &\scriptsize{$\alpha_{\rm opt,e}$\hspace{0.5cm}}		&		\scriptsize{$2.51\pm0.81$  (1588/50)\hspace{0.5cm} 2.50 }	&  \scriptsize{$ -$\hspace{0.5cm}}		\\
\scriptsize{Decay slope (before break)} &\scriptsize{$\alpha_{\rm opt,bb}$\hspace{0.5cm}}		&		\scriptsize{$ - $\hspace{0.5cm}  }	& \scriptsize{$0.921\pm0.163$ (36.9/28) \hspace{0.5cm}   1.05}	\\
\scriptsize{Decay slope (after break)}&\scriptsize{$\alpha_{\rm opt,ab}$\hspace{0.5cm}}		&		\scriptsize{$ - $\hspace{0.5cm}} & \scriptsize{$2.036\pm0.521$  (8.91/6) \hspace{0.7cm}   2.40 }			\\
\scriptsize{Break time (s)}&\scriptsize{$t_{\rm o, br}$ \hspace{0.5cm}}			&			\scriptsize{$ -$\hspace{0.5cm}}	&	\scriptsize{$1.71 \times 10^6$}\\
\scriptsize{Late spectral slope}    &\scriptsize{$\beta_{\rm opt}$\hspace{0.5cm}}			 &			\scriptsize{$-$\hspace{0.5cm}} &\scriptsize{$0.71\pm0.12$ (156.5/112) \hspace{0.7cm}   0.70}		\\

\\
\hline
\end{tabular}
\end{center}
\begin{center}
\begin{center}
\scriptsize{\textbf{Table 2. Parameters found of the early and late afterglow. }}
\end{center}
\begin{tabular}{ l c c c c c}
 \hline
 \hline
Early afterglow& & & & Late afterglow\\
\hline
\scriptsize{$\epsilon_{B,r}$}    & \scriptsize{ 0.40 $\pm$ 0.04} & &  & \scriptsize{$\epsilon_{B,f}$}    & \scriptsize{ $(1.1\pm0.1) \times 10^{-6}$ }  \\
\scriptsize{$\epsilon_{e,r}$}    & \scriptsize{0.45 $\pm$ 0.05} & & & \scriptsize{$\epsilon_{e,f}$}    & \scriptsize{0.45 $\pm$ 0.05}  \\
\scriptsize{$A_\star\,\,$}   & \scriptsize{$0.20\pm0.02$} & & & \scriptsize{$n\,\, (\rm cm^{-3})$}    & \scriptsize{10.0 $\pm$ 0.1}  \\
 \hline
\end{tabular}
\end{center}
\begin{center}
\end{center}
\begin{center}
\begin{center}
\scriptsize{\textbf{Table 3.   Quantities derived with our leptonic model for the early and late afterglow.  Quantities for the early afterglow are calculated\\ at $150$ s and for the late afterglow  at $5\times 10^4$ s. 
At the jet-break time ($ 1.6\times 10^6$ s), the jet opening angle and the bulk Lorentz factor\\ are $8.3^\circ$ and $6.9$, respectively.)}}\\
\end{center}
\begin{tabular}{ l c c c c c}
   \hline
 \hline
Early afterglow & & & & Late afterglow\\
\hline\hline
%
%
\scriptsize{$\Gamma_c$} & \scriptsize{397.5 $\pm$ 26.9} & & &   \\
\scriptsize{$B_r$ (G)} &  \scriptsize{123.9 $\pm$ 58.3} & & & \scriptsize{$B_f$ (G)} &\scriptsize{$(2.1 \pm 0.3) \times 10^{-1} $}  \\
\scriptsize{$\Gamma$} & \scriptsize{500}  & & & \scriptsize{$\Gamma$} &  \scriptsize{25.2 $\pm$ 7.8} \\
&  & & &  &  \\
\hline
\small{Synchrotron}\\
\hline
\scriptsize{$E^{\rm syn}_{\rm a,r}$ (eV)}    & \scriptsize{$(1.3\pm 0.4) \times 10^{-9}$} & & & \scriptsize{$E^{\rm syn}_{\rm a,f}$ (eV)}    & \scriptsize{$(7.7\pm 3.1)\times 10^{-5}$}  \\
\scriptsize{$E^{\rm syn}_{\rm m,r}$ (eV)}    & \scriptsize{2.1 $\pm$ 0.5 } & & & \scriptsize{$E^{\rm syn}_{\rm m,f}$ (eV)}    & \scriptsize{0.4$ \pm$ 0.1}  \\
\scriptsize{$E^{\rm syn}_{\rm c,r}$ (eV)}    & \scriptsize{$(2.5 \pm  0.7) \times 10^{-5}$} & & & \scriptsize{$E^{\rm syn}_{\rm c,f} $} (keV)   & \scriptsize{125.8 $\pm$ 32.6}  \\
%
\hline
\small{SSC}\\
\hline
\scriptsize{$E^{ssc}_{\rm m,r}$ (MeV)}    & \scriptsize{$90.5\pm 26.9$} & & &    &   \\
\scriptsize{$E^{ssc}_{\rm c,r}$ (keV)}    & \scriptsize{$(1.0\pm 0.1)\times10^{-3}$} & & &  &   \\
\hline
\end{tabular}
\end{center}
\end{document}